\documentclass[sigconf]{acmart}
\AtBeginDocument{%
  }

\copyrightyear{2025}
\acmYear{2025}
\setcopyright{cc}
\setcctype{by}

\acmConference[CHI '25]{CHI Conference on Human Factors in Computing Systems}{April 26-May 1, 2025}{Yokohama, Japan}
\acmBooktitle{CHI Conference on Human Factors in Computing Systems (CHI '25), April 26-May 1, 2025, Yokohama, Japan}\acmDOI{10.1145/3706598.3713244}
\acmISBN{979-8-4007-1394-1/25/04}

\usepackage{float}

\begin{document}

\title{Understanding Adolescents’ Perceptions of Benefits and Risks in Health AI Technologies through Design Fiction}

\author{Jamie Lee}
\affiliation{%
  \institution{Informatics}
  \institution{University of California, Irvine}
  \city{Irvine}
  \state{California}
  \country{USA}
}

\author{Kyuha Jung}
\affiliation{%
  \institution{Informatics}
  \institution{University of California, Irvine}
  \city{Irvine}
  \state{California}
  \country{USA}
}

\author{Erin Gregg Newman}
\affiliation{%
  \institution{School of Medicine}
  \institution{University of California, Irvine}
  \city{Orange}
  \state{California}
  \country{United States}}

\author{Emilie Chow}
\affiliation{%
  \institution{School of Medicine}
  \institution{University of California, Irvine}
  \city{Orange}
  \state{California}
  \country{United States}}

\author{Yunan Chen}
\affiliation{%
 \institution{Informatics}
 \institution{University of California, Irvine}
 \city{Irvine}
 \state{California}
 \country{USA}}

\renewcommand{\shortauthors}{Lee et al.}

\begin{abstract}
Despite the growing research on users’ perceptions of health AI, adolescents’ perspectives remain underexplored. This study explores adolescents’ perceived benefits and risks of health AI technologies in clinical and personal health settings. Employing Design Fiction, we conducted interviews with 16 adolescents (aged 13-17) using four fictional design scenarios that represent current and future health AI technologies as probes. Our findings revealed that with a positive yet cautious attitude, adolescents envision unique benefits and risks specific to their age group. While health AI technologies were seen as valuable learning resources, they also raised concerns about confidentiality with their parents. Additionally, we identified several factors, such as severity of health conditions and previous experience with AI, influencing their perceptions of trust and privacy in health AI. We explore how these insights can inform the future design of health AI technologies to support learning, engagement, and trust as adolescents navigate their healthcare journey. 
\end{abstract}



\received{12 September 2024}
\received[revised]{10 December 2024}
\received[accepted]{16 January 2025}

\maketitle

\section{INTRODUCTION}
With the rapid advancement of artificial intelligence (AI) and machine learning (ML), AI-driven technologies have surged across various contexts for health \cite{schwalbe_artificial_2020, jiang_artificial_2017, stromel_narrating_2024, berube_voice-based_2021, grzybowski_artificial_2020, tran_deep_2021}. According to the World Health Organization, health is defined as “a state of complete physical, mental and social well-being and not merely the absence of disease or infirmity” \cite{who_who_nodate}. Taking a holistic view, health experiences are multidimensional, encompassing medical care as well as personal health management in everyday contexts \cite{schramme_health_2023}. Consequently, AI-integrated applications for health (referred to as \textit{health AI} in this paper) hold great promise in transforming people’s health experiences, either as individuals directly interacting with AI-driven personal health technologies (e.g., tracking \cite{stromel_narrating_2024, davergne_wearable_2021, forman_ontrack_2019}, chatbots \cite{berube_voice-based_2021, luxton_ethical_2020, tosti_using_2024}) or as patients benefiting from healthcare providers who use an AI system (e.g., screening \cite{grzybowski_artificial_2020, freeman_use_2021, mitsala_artificial_2021}, treatment selection \cite{katzman_deepsurv_2018, huynh_artificial_2020, tran_deep_2021}). Health AI applications offer widespread advantages, such as personalized suggestions \cite{schwalbe_artificial_2020, rowe_artificial_2020, rajpurkar_ai_2022} and accessibility to mental health support \cite{tudor_car_conversational_2020, luxton_ethical_2020} to healthcare consumers, and higher diagnosis accuracy \cite{dilsizian_artificial_2014, nichols_machine_2019, huang_artificial_2020} and more efficient documentation \cite{falcetta_automatic_2023, lin_reimagining_2018, kocaballi_envisioning_2020} to clinicians, which ultimately benefit the patients being treated as well. 

Amid growing excitement for AI’s potential in healthcare, the field of Human-Computer Interaction (HCI) has long recognized that without user acceptance, new technologies may fail to achieve their intended impact. While promising, health AI technologies bring challenges, such as misclassification resulting from demographically biased training data, further perpetuating health inequities across various populations \cite{murdoch_privacy_2021, agarwal_addressing_2023}. Moreover, as the use of large language models (LLMs) is becoming prevalent, hallucinations (generated inaccurate information that seems sensible) in their responses can pose risks to users seeking medical advice \cite{bhattacharyya_high_2023, agarwal_medhalu_2024}. Many studies thus examined users’ perceptions, attitudes, and concerns regarding AI systems \cite{fast_long-term_2017, luca_liehner_perceptions_2023, beets_surveying_2023, cai_human-centered_2019, beede_human-centered_2020, wang_brilliant_2021}, uncovering considerations that shape human interaction with AI in health settings. One key factor affecting perception and usage of AI is trust, which reveals deep complexity as on one hand, some studies find that clinicians place less trust in AI systems due to its black box nature \cite{wang_brilliant_2021}, calling for explanations to increase trust. On the other hand, explanations could increase trust too much, leading to overreliance from novice clinicians \cite{gaube_as_2021}. Providers and patients also have voiced concerns about privacy and security, potentially leading to resistance against AI \cite{beets_surveying_2023, young_patient_2021, dilsizian_artificial_2014}. Additionally, human-AI collaboration has served as a framework for clinicians to safely optimize AI’s utility, ultimately enhancing health outcomes on behalf of the patients \cite{cai_human-centered_2019, beede_human-centered_2020, asan_artificial_2020, wang_brilliant_2021}. Likewise, understanding how users perceive AI’s benefits and concerns has been ongoing for designing and implementing health AI technology for various health experiences, from clinical to personal health contexts. 

Not surprisingly, existing literature on health AI technologies has predominantly centered on the perspectives of clinicians, finding that although they feel optimistic, they face technical challenges and trust issues during adoption \cite{cai_human-centered_2019, beede_human-centered_2020, asan_artificial_2020, wang_brilliant_2021}. Recently, HCI research has begun exploring the perspectives of other stakeholders such as adult patients \cite{you_beyond_2023, kim_how_2024, yoo_patient_2024, figueiredo_powered_2023}, parents \cite{ramgopal_parental_2023, petsolari_socio-technical_2024}, and caregivers \cite{li_personalized_2020, mccradden_ethical_2020, f_corbett_virtual_2021} in both clinical and personal health settings. For example, Yoo et al. investigated patients’ perceptions of AI-based predictions in the context of schizophrenia relapse and found that while patients recognized the potential benefit of self-reflection through the AI outputs, they felt that the AI model's definition of relapse did not accurately reflect their lived experiences, negatively affecting their trust \cite{yoo_patient_2024}. Across different populations, common themes include the perceived benefits of AI for health, concerns about trust and privacy, and the emphasis on human-AI collaboration. 

Due to the primary focus on adult stakeholders (e.g., clinicians, patients, and parents), there is a gap in the literature, with adolescents’ perspectives on health AI remaining underexplored. Adolescence is a critical stage for establishing lifelong health behaviors, as physical and mental health issues that can possibly affect adulthood begin to emerge \cite{holmbeck_developmental_2002, catalano_worldwide_2012}. During this time, adolescents start to gain autonomy in managing their health \cite{reiss_health_2005, cha_transitioning_2022, su_creating_2024, hong_care_2016, ankrah_plan_2023, huang_transition_2011, zehrung_transitioning_2024}, often utilizing technologies such as the internet and social media for accessing health information \cite{gray_health_2005, freeman_how_2023}. With the growing trend of health AI, adolescents will most likely be exposed to more AI applications in health, along with the potential benefits and risks that could impact them during this developmental stage \cite{rowe_artificial_2020, hoodbhoy_machine_2021}. These particular considerations underscore the importance of including their voices in envisioning the future of health AI. 

To address this gap, we utilized Design Fiction to investigate the potential benefits and risks surrounding health AI technologies from the perspective of adolescents \cite{dunne_speculative_2013, blythe_research_2014}. Design Fiction methods have been widely used in HCI to explore possible futures to gain deeper insights into the present, as well as facilitate conversation about the kind of future people desire \cite{zimmerman_role_2008, dunne_speculative_2013, blythe_research_2014, linehan_alternate_2014, dourish_resistance_2014}. This method is particularly effective to study adolescents’ perspectives on health AI, as it can surface concerns and aspirations before they fully adopt and are impacted by such technologies. Through Design Fiction methods, we sought to answer the following research questions: 

\begin{enumerate}
    \item \textit{RQ1: How do adolescents perceive the benefits and risks of health AI technologies for clinical care and personal health management?}
    \item \textit{RQ2: What factors influence adolescents’ perception of the benefits and risks of health AI technologies?}
\end{enumerate}

To answer these questions, we collaborated with pediatricians to create current and future-oriented design fictions as probes to gain a holistic understanding of adolescents' perceptions regarding health AI technologies. Grounded in existing literature and expert input, we designed four fictional scenarios illustrating potential AI integration in clinical and personal health contexts. Using these scenarios, we conducted semi-structured interviews with 16 adolescents aged 13-17 to elicit dialogue about their attitudes, perceived benefits, and concerns about AI utilization in health technologies. 

Our findings reveal that adolescents view health AI with cautious optimism, recognizing both benefits and risks specific to their age. On one hand, adolescents saw health AI as beneficial for learning health skills. On the other hand, they expressed apprehension about health AI technologies sharing private information with their parents. Overall, adolescents preferred human involvement when utilizing health AI and wanted to collaborate with both their providers and AI in decision-making. By comparing adolescents' perceptions across four clinical and personal health scenarios, our study identified key factors shaping trust, including complex health conditions, empathy, and prior encounters with AI. Privacy concerns varied based on medical and personal data sensitivity and AI data collection methods. Based on these findings, we discuss ways to reflect on these insights for future designs of health AI technologies.

We make the following contributions to the HCI health communities: 

\begin{enumerate}
    \item We provide in-depth descriptions of adolescents’ perceptions about the imagined uses, benefits, and concerns for four different health AI technologies used for both clinical care and personal health management.
    \item We provide new insights into how this particular age group perceives health AI and suggest design considerations to accommodate their unique needs.
    \item We introduce additional factors, such as complexity of health conditions, lack of empathy, and previous experiences with AI, that influence their perceptions of trust in health AI. 
    \item We offer nuanced understandings of how adolescents’ perceptions of data privacy concerns are shaped by factors including type of sensitive data and type of additional data being recorded by AI.
\end{enumerate}

\section{RELATED WORK}
\subsection{Health AI Technologies}
Research on AI-integrated systems has grown, focusing on assisting clinicians in decision-making through the implementation of classical ML (e.g., predictive modeling) and deep learning (e.g., neural networks) \cite{middleton_clinical_2016, lee_human-ai_2021, busnatu_clinical_2022}. These AI systems analyze vast datasets to support doctors in disease diagnosis, risk analysis, treatment selection, and data management \cite{magrabi_artificial_2019, sutton_overview_2020, yin_role_2021, tierney_ambient_2024}. For instance, Zhang et al. developed an AI-powered platform that predicts sepsis development \cite{zhang_rethinking_2024}. However, despite clinicians being receptive to AI support for medical decision-making regarding sepsis, they felt that the current AI platform was not only ineffective but also added additional unnecessary work. Furthermore, predictive algorithms have the potential to widen existing health disparities. Obermeyer et al. found that health data are often labeled with errors reflecting structural inequalities, leading the algorithm to misclassify sick Black patients as similar to healthy White patients and thereby reducing their access to healthcare resources \cite{obermeyer_dissecting_2019}. Other recent systems, such as ambient scribes, have been developed to focus on alleviating clinical documentation by emulating transcription and summarization capabilities through NLP techniques \cite{coiera_digital_2018, lin_reimagining_2018}. However, less work has addressed the technical barriers to implementing AI scribes, such as the need for these tools to adapt to diverse populations, accommodating differences in languages, styles, accents, and visit types \cite{ghatnekar_digital_2021}.

There has also been a rapid growth of AI-powered personal health technologies including mobile health apps, wearable, fitness tracker, conversational chatbots that utilize AI and ML techniques such as NLPs and LLMs \cite{haque_overview_2023, stromel_narrating_2024, g_mitchell_reflection_2021, tudor_car_conversational_2020, milne-ives_effectiveness_2020, you_user_2022, lee_i_2020}. These AI-driven technologies aim to improve health outcomes by supporting people to self-manage and improve their health through personalization and recommendations. For example, Mitchell et al. found that personalized goal recommendations for nutrition intake led to self-reflection and learning \cite{g_mitchell_reflection_2021}. Another popular personal health technology is AI-based chatbots which can provide easier access to mental health services such as cognitive behavioral therapy, help alleviate symptoms, and promote disclosure \cite{tosti_using_2024, inkster_empathy-driven_2018, fitzpatrick_delivering_2017, sweeney_can_2021}. 

Although prior work on health AI technologies has been extensively explored, fewer studies have examined these technologies across both clinical and consumer contexts. As the development of health AI technologies carries strong prospects for enhancing health management practices, their effectiveness ultimately depends on the stakeholders and their perspectives on AI's impact across diverse health contexts. Understanding the perspectives of users regarding these tools is key to realizing the full potential of AI in health management. 

\subsection{Perceptions, Benefits, and Concerns of AI in Health}
Past studies have revealed that the public often views AI as a double-edged sword, viewing both risks and benefits simultaneously \cite{kelley_exciting_2021, schwesig_using_2023}. In a survey across eight countries by Kelley et al. \cite{kelley_exciting_2021}, AI was believed to be transformative, especially in healthcare, and generally viewed as “exciting”, “useful”, “worrying”, and “futuristic”. Despite the overall optimistic attitude towards AI, several factors such as context, trust, privacy, and autonomy were found to significantly influence the willingness to use AI \cite{magrabi_artificial_2019, asan_artificial_2020, esmaeilzadeh_patients_2021}.

Studies have revealed conflicting views on the trustworthiness of AI systems compared to human decision-making, leading to nuanced perceptions about trust \cite{larosa_impacts_2018, beede_human-centered_2020, asan_artificial_2020, chen_explainable_2022, araujo_ai_2020, fogliato_who_2022, lee_who_2021}. Factors such as explainability, transparency, interpretability, usability, and education can significantly impact a healthcare provider’s trust in medical AI and therefore, the decision-making process \cite{tucci_factors_2022}. Cai et al. conducted an interview study with pathologists newly introduced to an AI assistant in diagnosing prostate cancer and found that they sought information about the AI beyond explainability, such as its functionality and medical perspective, to foster trust and collaboration \cite{cai_hello_2019}. A review by Steerling et al. identified three overarching themes that influence trust in AI within healthcare settings–individual characteristics, AI characteristics, and contextual characteristics \cite{steerling_implementing_2023}. While most studies focused on how individual or AI characteristics shape trust in AI, the authors highlight the complexity and contextual nature of trust, emphasizing the importance of adopting a holistic view to understand the concept.

The autonomy of AI–referring to the degree to which an AI system can make decisions without human intervention–was another factor that influenced people’s perceptions. For example, a study by Kim et al. explores how the level of AI autonomy could affect patients’ perspectives in medical decision-making in the context of pregnancy care \cite{kim_how_2024}. Patients preferred a human provider to be involved across all levels of AI autonomy. Similarly, Esmaeilzadeh et al. found that individuals with chronic illnesses may not trust AI clinical applications if human providers are not involved at all \cite{esmaeilzadeh_patients_2021}. Most studies reveal users’ resistance to fully autonomous AI systems in healthcare settings as they still prefer human involvement. 

Data privacy concerns about health AI are well documented \cite{wu_public_2023, chew_perceptions_2022, beets_surveying_2023}, as AI introduces new challenges, including the need for vast quantities of data, unauthorized third-party data access, heightened risks of data breaches, and dissemination of realistic spurious images of people (e.g., deepfakes) \cite{murdoch_privacy_2021, khalid_privacy-preserving_2023, williamson_balancing_2024, lee_deepfakes_2024}. Several contextual variables affect individuals' perceptions of such risks. For example, when comparing acute and chronic conditions, participants with acute illnesses reported significantly greater privacy concerns when interacting with AI applications. In contrast, those with chronic conditions showed no notable difference in privacy concerns between AI-assisted interactions and traditional provider visits \cite{esmaeilzadeh_patients_2021}. In the context of mental health, individuals have raised data privacy concerns about emotion AI–or technologies attempting to automatically infer human emotion and act accordingly–reporting the potential violation of privacy \cite{roemmich_emotion_2024}. 

These studies highlight the multitude of factors, such as individual characteristics, human involvement, and privacy, that affect AI perceptions, ultimately influencing trust and usage. This underscores the necessity to understand patients' perspectives alongside those of clinicians \cite{dunbar_my_2022}. However, HCI research predominantly focuses on adults, particularly clinicians and adult patients when examining health AI. As adolescents navigate the most technologically advanced period in history, they will most likely emerge as primary users of AI, yet their representation in research remains limited. To address this gap, we aim to explore adolescents’ perceptions, benefits, and concerns about health AI.
 
\subsection{HCI Technologies for Adolescents}
Adolescence is a pivotal stage in life, marked by significant physical, cognitive, and social changes \cite{frech_healthy_2012, holmbeck_developmental_2002}. To achieve and maintain good health, adolescents require age-appropriate education and opportunities to develop essential life skills, and guidance from their support systems. Recent HCI research on adolescents and health technologies show great promise in supporting their health management skills \cite{hong_care_2016, zehrung_investigating_2021, ankrah_plan_2023, rahman_adolescentbot_2021, freeman_putting_2023, su_creating_2024}. This promise lies in the ability of these technologies to support adolescents in active engagement in their health \cite{hong_care_2016, ankrah_plan_2023}, tailored health education \cite{freeman_putting_2023}, and interactive tools for self-management \cite{rahman_adolescentbot_2021, freeman_putting_2023}. For example, Freeman et al. hosted workshops with teenagers to co-design a resource to support positive tracking experiences \cite{freeman_putting_2023}. The participants were thoughtful and wary about the negative aspects of self-tracking, yet they also recognized the value, ultimately expressing a desire for tools that accommodate the diverse experiences of teenagers. Additionally, adolescents may begin to co-manage their health with their parents at this age \cite{toscos_best_2012, pina_personal_2017, lee_mobile_2023, su_data-driven_2024, su_creating_2024}. While co-management can promote healthy habits, it can also create tensions, as sharing too much data may raise privacy concerns \cite{pina_personal_2017}.

However, despite the growing research on various support systems, adolescents often encounter unique barriers, such as stigma and confidentiality concerns, when seeking help for health issues \cite{carlisle_concerns_2006, fuentes_adolescents_2018, michaud_assessing_2015, su_hidden_2024}. Adolescents may fear being judged if they seek help, particularly for sensitive issues such as mental health, sexual health, substance abuse, or even serious illnesses, and thus, having a way to discuss these issues with anonymity could be especially beneficial \cite{lucas_reporting_2017}. Adolescents also often have nuanced concerns about privacy, particularly when it comes to sharing sensitive information with healthcare providers. Some fear that personal details will be disclosed to their parents, while others desire greater control over what information to share \cite{bergman_teen_2008, hagstrom_views_2022}. For instance, in Bergman et al.’s study, while teenagers were enthusiastic about the concept of a teen portal, they expressed strong desires to keep certain billing information private from their parents \cite{bergman_teen_2008}, underscoring their need for both active health management and control over their privacy.

We aim to investigate adolescents' perceptions about health AI technologies to add to existing literature and provide guidance on developing youth-centered solutions that address these concerns and foster health management skills as they transition into adulthood.

\subsection{Design Fiction}
To examine adolescents’ perspectives and concerns about health AI technologies, we utilized Design Fiction, a methodology that creates fictional scenarios, narratives, or prototypes to explore speculative futures \cite{zimmerman_role_2008, dunne_speculative_2013, blythe_research_2014, linehan_alternate_2014, dourish_resistance_2014, lindley_back_2015}. Design-oriented research is fundamentally future-focused, rooted in exploring alternative possibilities made possible by technological progress \cite{sterling_cover_2009, noortman_hawkeye_2019, aki_tamashiro_introducing_2021, petsolari_socio-technical_2024}. The goal is to think critically about the future, questioning assumptions and challenging expectations about the evolution of technologies and their societal impact \cite{wong_real-fictional_2017, venta-olkkonen_nowhere_2021}. Design fiction probes are crafted to elicit open-ended responses from participants about the values, possibilities, and limitations of emerging technologies \cite{schulte_homes_2016, rubegni_dont_2022, rezwana_user_2023, lallemand_trinity_2024}. Drawing from these design approaches, our objective is to utilize Design Fiction to not only probe their imagined uses but also to understand what shapes their understanding about health AI.  

\begin{figure*}[!h]
  \centering
  \includegraphics[width=15.5cm]{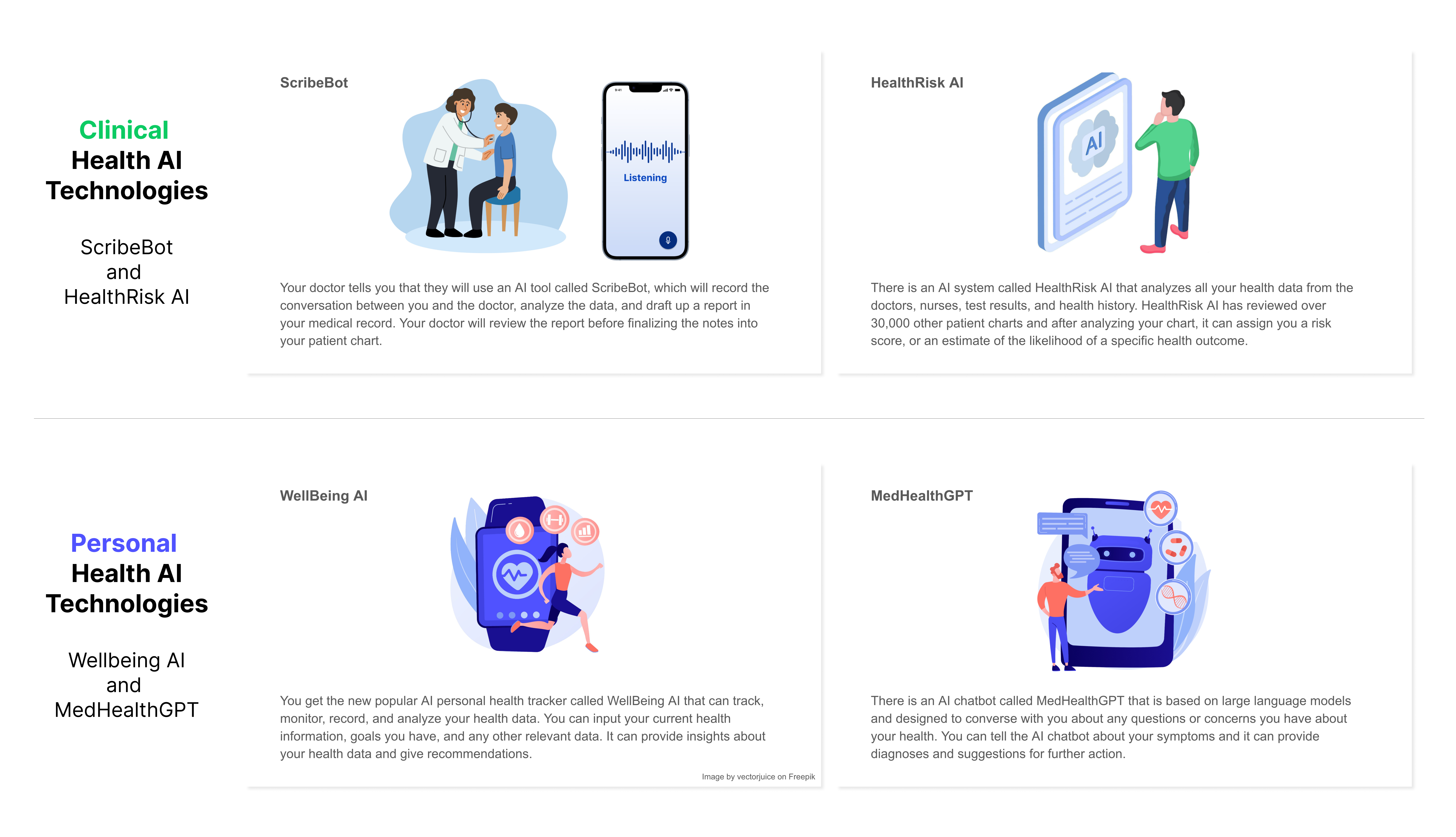}
  \caption{Scenarios of health AI technologies developed and used for interviews}
  \label{tab:figure1}
  \Description{The image is divided into four equal parts with each part displaying either a clinical or personal health AI technology. The upper left corner contains the AI digital scribe scenario. It has a vector illustration image of a doctor interacting with a patient by taking his heartbeat. On the left of the pair, there is a phone with the text 'Listening' and a record button. The upper right corner contains the AI risk assessment tool scenario. There is a vector illustration image of a person standing in front of a very large display with the word 'AI' on it and he is looking at the screen. The bottom left corner contains the AI personal health tracker scenario. There is a vector illustration image of a large smartwatch with icons that represent water, exercise, and metrics floating on top of it. A smaller scaled lady is running in front and past the smartwatch. The bottom right corner contains the AI chatbot scenario. There is a vector illustration image of a very large tablet with a robot appearing on the screen. Icons of a heartbeat, medication pills, and DNA are floating on the upper right of the tablet. A man is on the left of the tablet, speaking to it, as indicated by speech bubbles above them. For each scenario, there is a paragraph of text underneath the image that explain the situation.}
\end{figure*}

\section{METHODS}
This exploratory study seeks to gain a deeper understanding of adolescents’ perceptions, concerns, and attitudes towards health AI technologies. We first developed the scenarios to represent a variety of health AI tools and then conducted in-depth interviews with adolescents to gather nuanced insights into their views and expectations. 

\subsection{Scenario Development}
To design scenarios that effectively represent the field of health AI technologies, it was crucial to select tools that are both relevant and impactful to society. The scenario development involved all authors, including two domain experts: a clinical professor and an associate clinical professor of Internal Medicine and Pediatrics. 

We began by conducting a broad literature search to identify key health AI technologies \cite{jiang_artificial_2017, davenport_potential_2019, schwalbe_artificial_2020, abbasgholizadeh_rahimi_application_2022, yin_role_2021, busnatu_clinical_2022, iqbal_clinical_2021, alowais_revolutionizing_2023}. Although not a systematic review, we used scenario development methods similar to those in \cite{petsolari_socio-technical_2024, schulte_homes_2016, rubegni_dont_2022} and sought input from the expert doctors to identify health AI technologies most relevant to adolescents. We found AI integration into two main aspects of healthcare: primary care and personal health management. In the clinical space, the most common cases of AI utilization were disease diagnosis, risk assessment, treatment, and data management. AI techniques, such as machine learning and natural language processing, have been used in many aspects of the clinician’s workflow, either making predictions or extracting information from medical data (e.g., clinical notes). For personal health management, technologies integrating AI included mobile health (mHealth) apps, wearables, smartwatches, and conversational chatbots. Given the growing popularity of personal tracking technologies \cite{choe_understanding_2014, epstein_mapping_2020, oygur_lived_2021, freeman_challenge_nodate, silva_unpacking_2023} and conversational agents \cite{kim_can_2018, tudor_car_conversational_2020, may_conversational_2024, lopatovska_designing_2023}, we selected these tools to ensure our scenarios reflect the current and future landscape. Based on our literature review, we chose two scenarios from the clinical domain and two from the personal health domain. 

During this time, we began the initial drafts of the fictional scenarios for each of the health AI tools. Each scenario consisted of 1) a general overview of the current possible uses of the technology and 2) a description of its advanced functionalities and potential future applications (See Table 1). We iteratively developed the scenarios with the two expert doctors providing frequent insights into the practical challenges and opportunities within the field. To present the scenarios clearly, we designed the narratives using visual examples and supplementary text. For each scenario, an image representing the health AI technology was shown alongside a few sentences describing the situation the image presented (See Figure 1). We gathered the illustrations from websites that offered free stock images and vectors and constructed the scenarios using Figma and Google Slides.

\subsection{Recruitment and Data Collection}
With IRB approval, we recruited English-speaking adolescent participants (age 13 to 17) from two high school summer programs designed to inform them about careers in healthcare. The programs were conducted at an academic institution in the western United States. Recruitment materials were handed to attendees and emailed after the programs ended. We also utilized snowball sampling recruitment. Interested participants filled out a pre-screening questionnaire to ensure eligibility. Both parents/guardians and participants signed the consent/assent forms, and the authors received them via email or in person. The consent/assent forms were reviewed at the beginning of each interview. 16 people filled out the screening questionnaire, signed the consent forms and completed the interview from June–August 2024. There were 14 female and 2 male participants, and although their ages ranged from 13-17 years old, the average age was 16 years old.

We conducted semi-structured interviews using the fictional scenarios as probes to encourage open responses from participants. The interview started with general questions about the participants' demographics and their overall understanding and beliefs about health. Following this, we explored participants’ general thoughts and knowledge of AI. Once this context was established, we transitioned to the fictional scenarios.

For each scenario, we presented the design fiction probes and described the AI technology, its purpose, and how it could be used. The interviewer shared their screen to show the probe. For instance, when discussing the ScribeBot scenario, the Google Slide containing the image and description of the ScribeBot was shown. We first discussed the overview of the current possible uses of the technology and inquired about their perspectives. For example, we probed with questions such as “\textit{What are your initial thoughts about this tool?}” or “\textit{What are some advantages of using this tool?}” or “\textit{Do you have any concerns about [AI tool name]? If so, what are they?}” Afterward, we introduced the advanced features and continued to probe their thoughts. We asked about their envisioned ideas about the tool by asking “\textit{How would you imagine using this tool?}” or “\textit{In what situations do you imagine this tool being helpful for you?}” We emphasized that there are no right or wrong answers as young participants may be more susceptible to social desirability bias. We alternated the order of scenario presentation to minimize order effects and reduce potential bias. The one hour interviews were held over Zoom. They were audio recorded and transcribed by the first author. Each participant received a \$25 Amazon gift card as compensation.

\begin{table*}[ht]
\footnotesize
\centering
\resizebox{\textwidth}{!}{%
\begin{tabular}{p{4cm}p{5cm}p{5cm}}
\toprule
\textbf{Health AI Technology} & \textbf{Description of Current Concept} & \textbf{Description of Future Concept} \\ 
\midrule
ScribeBot (clinical) & \textbf{An AI ambient scribe} is set up to transcribe and summarize the conversation between the patient and doctor, extracting the most relevant information. & Using cameras to capture face and body movements and utilizing AI to analyze the recordings, potentially providing better analysis and insights about the patient.\\
\\

HealthRisk AI (clinical) & \textbf{An AI clinical prediction tool} uses large datasets and decision models to analyze a patient’s characteristics and calculate a risk score for health outcomes or diseases. & Using more advanced AI models to recommend medications for health conditions based on risk scores and directly prescribe medication without doctor’s approval. \\
\\

Wellbeing AI (personal) & \textbf{An AI health tracker} that supports self-tracking and analyzes comprehensive tracking data to offer personalized insights and recommendations on health management. & Using AI facial tracking to analyze emotions as well. It can provide instant feedback and send any health data to your doctor to alert them when it predicts a serious upcoming health issue. \\
\\

MedHealthGPT (personal) & \textbf{An AI chatbot} designed to mimic human conversation and support functions including suggesting possible health conditions, providing solutions, and offering emotional support. & Using human-like emotional intelligence and acting as a therapist, MedHealthGPT can diagnose mental health issues, engage in meaningful, human-like conversations, and provide mental health therapy sessions. \\
\bottomrule
\end{tabular}%
} 
\caption{Current and future concept descriptions of health AI technologies discussed in interviews}
\label{tab:my-table}
\end{table*}

\subsection{Data Analysis}
The first author began data analysis after the first two interviews and continued throughout data collection. Qualitative memos were utilized to foster analytical thinking during data collection and analysis. After we reached data saturation, or when the interviews no longer brought new insights, the first and last authors conducted thematic analysis using open coding \cite{braun_using_2006}. Taking an inductive, bottom-up approach, they derived codes from the data. After discussing initial codes, the first author coded the remaining transcripts, meeting regularly with the second and last author to refine codes. Key codes included “human involvement,” “lack human connection,” and “family health history”, focusing on factors that influenced participants’ views. The first author used affinity diagrams to organize codes into broad themes. We revised the codebook over several weeks. Weekly meetings with the first, second, and last authors continued until codes and themes were finalized, identifying trends in perceived benefits, concerns, and imagined AI uses.

\section{FINDINGS}
In this section, we discuss findings from our analysis of the interview transcripts. We sought to use scenarios to probe a variety of possible imaginaries involving AI and health with adolescents, identifying broad attitudes about health AI technologies in the process. We describe an overview of how participants generally responded to each scenario, focusing on their perceived benefits, concerns, and envisioned future uses of health AI technologies. The analysis revealed that the overall tone of the adolescents stayed positive but cautious. 

\subsection{Clinical Scenario One: ScribeBot}
The first clinical scenario elicited adolescent perspectives about a support tool called ScribeBot, which transcribes and summarizes consultations between a patient and their doctor. 

\subsubsection{Potential Benefits of ScribeBot: Helpful and Complementary to the Doctor's Role}
Intended to be used by doctors during general medical exams, ScribeBot was viewed positively among the majority of the adolescents, who often described it as “\textit{good} ” (A01) and “\textit{helpful}” (A07). For example, A01 considered, "\textit{I think it's efficient. And since it is like AI, it can just process the information really quick.}” Some participants expressed admiration, as A07 mentioned: “\textit{I think this tool is pretty genius. I feel like it definitely makes the doctor’s job much easier.}” They clearly saw the advantages of the tool for both doctors and patients. A11 described how the ScribeBot could potentially help the doctor “\textit{in case [he] misses anything while jotting down his notes. And if the AI summarizes the whole conversation and makes the report on its own, I think it would lessen the load off the doctors and this makes doctors' lives easier.}” They believed the ScribeBot would allow the doctor to pay more attention to patients, instead of taking notes. A16 “\textit{would rather have the doctor more focused on me then writing some stuff down.}” According to participants, the AI tool’s quick information processing could help doctors save time, allowing them to see more patients and provide more focused care. 

The positive reception of ScribeBot came with the belief that its role should complement the doctor’s. For instance, A05 maintained that “\textit{I think the doctor should obviously use [ScribeBot] as a resource because AI is meant to be a resource and for it to help. But I don't think the doctor should solely rely on it.}” The tool was seen as a useful aid for supporting the doctor, but not to be completely depended on. 

\subsubsection{Perceived Concerns of ScribeBot: Misinterpretation, Confidentiality Issues, and the Prescence of AI}
As the adolescents were probed more about certain situations that could occur using ScribeBot, they voiced concerns that prevented complete trust in the AI tool. First, participants worried about the likelihood for ScribeBot to misinterpret information during patient-provider conversations. A05 explained their thoughts: “\textit{If [ScribeBot] doesn't catch on what you're saying, words can get twisted or miscommunicated. And then it can take that information in a wrong way, and spit out wrong information.}” A07 pointed out the possibility that “\textit{maybe a patient has something that they're mildly concerned about, but the AI kind of thinks it's very severe and a really big deal, but then it's really not.}” According to A07, ScribeBot could potentially misunderstand the patient-provider conversation, resulting in faulty summaries. These reflections indicate that adolescents felt a human would make better inferences based on their conversation compared to an AI technology. Misinterpretation could lead to inaccurate analysis of the conversation, potentially leading to negative health consequences for the patient.

Second, adolescents expressed distinct confidentiality concerns regarding who could access their personal information from the AI tool. Most participants were uneasy about the AI tool potentially leaking their medical data, as A01 indicated apprehension about corporations misusing the exposed data: “\textit{Just the fact that other companies could maybe take that information and that recording and use it for their own purposes.}” It was unclear to the adolescents how the data breaches while using AI technology would occur though, as A03 noted, “\textit{I don’t know how that works but [hacking] could be a possibility.}” A04 also doubted the security of the AI tool handling sensitive personal information, stating: “\textit{I would not trust this device because I think, especially with technology, it can easily be hacked and I think information can be taken out of technology very easily. [...] I feel like there’s gonna be a point where AI becomes more intelligent than we are. [...] I don’t know what it could do with the information.}” This suggests that the participant believes AI technology will advance, but it could introduce greater security risks due to uncertainty surrounding data management. 

Interestingly, another confidentiality concern expressed was about sensitive information being shared with their parents rather than the public. A12 stated: “\textit{Sometimes you would talk to your doctor about mental health issues and if you were to do that, potentially the ScribeBot could write it down and then the doctor would let your parent know, and you didn’t want your parent to know, so it’s just letting your parent know what you didn’t want them to know.}” This comment indicates that the presence of an AI tool could prevent some adolescents from being fully honest during a medical visit, due to the fear that their parents might discover sensitive information without consent. 

In contrast, a few participants did not have data privacy concerns personally as they did not know how their own data could be taken advantage of. For example, A14 wondered, “\textit{I'm not really sure what anybody would want to do with my medical or personal health data other than just know everything about me. [...] I don't think anybody collecting the data would use it against me. I'm not really sure how they would use it against me.}” When asking A16 about having confidentiality concerns about their medical data, they responded, “\textit{Who would really care to look at it?}” For both participants, data breaches were not perceived as a threat, as they believed their health data was not useful enough to be exploited by someone. On the other hand, some participants were willing to accept certain privacy risks if they believed that AI would use their data to improve patient care. As A14 explained, “\textit{I don't really care who's collecting the data at the end of the day as long as it's helping me, so I think I would be comfortable to talk about any topic as long as I'm consenting to using the ScribeBot.}” 

Finally, there were situated concerns about ScribeBot listening in depending on the topics discussed. While adolescents were accepting of the AI tool’s presence during the general check-up exam, when asked specifically about consulting sensitive issues (e.g., mental health history, sexual health, and substance abuse), they shared their discomfort. For example, A01 explained, “\textit{I think I wouldn’t be comfortable just because those topics are more sensitive and more related to your mental health.}” A15 commented on how this tool could potentially hinder adolescents trying to open up about their struggles, “\textit{I know there’s a lot of teens that do illegal drugs [...] and I think it could be a really sensitive subject for them, especially if they're trying to get off of it. And it's already kind of hard to talk to a person face to face about the issues that they're dealing with, and so having an AI listen in would even make it even harder.}” Adolescents may already be reluctant to confide in their doctors about illegal activities due to the fear of incrimination, and the presence of an AI tool documenting them could become yet another barrier to getting the help they need.

\subsubsection{Future Concept of ScribeBot: Excessive and Invading Privacy}
When being asked about a future scenario where ScribeBot can capture the entire interaction on camera, the majority of the participants felt conflicted between the prospective benefits and their discomfort with recording. A few noted the potential advantages of more data for better analysis, as A09 contended, “\textit{I think that's definitely way better than just having [ScribeBot] hearing the conversation and stuff cause how I was saying before, it would definitely be able to pick up some type of body language now and some type of extra help to analyze the data and everything.}” However, most adolescents expressed how uneasy and unnecessary it would be. For example, A08 adamantly responded, “\textit{Okay, that is like a strict no. I don't think it necessary for you to record, like with a camera the entire thing, the entire visit.}” The risk of confidentiality breaches seemed higher when more identifiable information was involved, as A01 explained,“\textit{I feel like having both the listening aspect and the video–it just seems like a lot of information that, put in the wrong hands, could end up being bad... you can link the conversation being recorded so like an actual face.}” The idea of excessive recordings of one's conversation and face made participants feel extremely uncomfortable and worried that potential data breaches would result in worse outcomes as they could be more easily identified. A03 expressed even feeling vulnerable, “\textit{It may definitely be concerning to me because I’d feel unsafe in that environment, not because the AI could do anything bad but [...] you feel safer with a real person, right?}” A03 showed hesitation towards the AI technology and felt safer when humans were involved. These responses indicate how privacy risks are perceived to be higher and levels of trust are lowered regarding excessive monitoring and recording.

\subsection{Clinical Scenario Two: HealthRisk AI}
The second clinical scenario introduced a clinical decision support tool called HealthRisk AI that computes a risk score for health consequences based on a large database of patient charts. The concept focused on risk scores for both common illnesses (e.g. cold, flu) and chronic diseases (e.g. diabetes).

\subsubsection{Potential Benefits of HealthRisk AI: Supporting Preventative Action Based on Family Health History}
HealthRisk AI was considered generally helpful among adolescents. The risk score was generally perceived to be personalized and trustworthy due to the AI analysis based on “30,000 patient charts.” As A05 said, “\textit{Yeah, honestly, I would [trust the score] because it’s taking information from 30,000 other patients and because it’s able to take that information from other people and use that to help assess your score.}” Adolescents noted the value in the AI system analyzing and comparing other patient charts to theirs. Many imagined the doctors would use HealthRisk AI as additional assistance in patient care. A07 relayed the idea: “\textit{I think a doctor should try and make their own predictions first and then maybe cross-check with HealthRisk AI to see if they overlook something.}” Participants presumed that the doctors would share the risk scores with them to take preventative measures together. For instance, A11 envisioned, “\textit{The doctor would probably just feed it my health data and it would spit out a risk or estimate of my likelihood of a health outcome. And then doctor would probably provide me with those facts and say, ‘Here's this, here's that, and then here's something you can prevent from this to happen if your score is high.’}” This expected collaboration among patients, clinicians, and the AI tool suggests that adolescents want to be involved in taking preventative action based on the risk score. In fact, the majority of the participants had a family history of health conditions such as high cholesterol, heart issues, and even skin cancer, and they stated that knowing one’s health risks could lead to proactive steps. A09 described how it could personally be helpful: “\textit{[HealthRisk AI] would give me a pretty good advantage of knowing what could happen and everything, and how to prepare myself, or maybe how to even prevent that from happening.}” These adolescents were aware of their family health issues and saw the potential advantages of AI in supporting their health journey through education and collaboration with their doctor for preventative care. 

\subsubsection{Perceived Concerns of HealthRisk AI: Risks Associated with Varying Health Conditions and Lack of Human Insight}
Participants also discussed several issues regarding the risk scores. First, they had concerns about the perceived accuracy, which varied depending on the types and severity of health conditions. We first observed distinct differences in attitudes regarding physical and mental health conditions. Generally, AI predictions for physical health were seen as more reliable and accurate than those for mental health conditions as people “\textit{experience [mental health issues] on different levels}” (A01). A03 elaborated on the variability of mental health: “\textit{Yeah, definitely like mental health you can't measure as well as physical health. Everyone's benchmark for where they should be is gonna be different because everyone has different living conditions and how much stress they can tolerate.}” In this respect, adolescents thought risk scores for mental health would be less accurate because people experience stress uniquely, and the AI system would not be able to properly evaluate them. Regarding the severity of health issues, we also found that adolescents placed more trust in HealthRisk AI’s analysis of minor health conditions compared to serious or rare situations. When probing A11 about trusting the tool, they replied, “\textit{Yes, I would because if it's reviewed over 30,000 patient charts, it has an idea of common health concerns, the cold, or the flu. If it's used for common illnesses, then I would trust it. But if it's more rare or not so common health concern, then I wouldn't.}” This participant considered the AI tool to be less accurate for rare health conditions because they assumed the difficulties for the AI to be trained on a scarce number of rare cases.

Additionally, participants felt that the lack of personal context and human insight would negatively impact the perceived accuracy of HealthRisk AI. Many adolescents argued that the system lacked the knowledge and understanding of the patient’s life holistically, such as “\textit{knowing what neighborhood you live in, and knowing your family life, and where you work, and how much income you make}” which A10 felt was necessary for proper analysis. Participants believed that without understanding human experiences, the tool would produce less accurate analyses. A08 emphasized this point, stating they would trust their doctor over the AI because “\textit{[My doctor] knows me at a human level [...] With the HealthRisk AI, I fear that it’s solely mathematical. [...] I just can’t see it understanding my lifestyle to that extent to be able to give me a number.}” In this situation, it was difficult to fathom how the AI system could understand the participants on a deeper level like their human providers could. Since an AI does not have its personal context nor humanlike understanding, adolescents felt it could be less accurate and, therefore less trustworthy compared to a human doctor.

\subsubsection{Future Concept of HealthRisk AI: Resistance to Full AI Autonomy}
In discussions about the future concept of HealthRisk AI prescribing medications without a doctor’s approval, adolescents highlighted reservations about the lack of doctor oversight in the prescription functionality. When asked about this situation, A04 responded, “\textit{But I don't think I personally would [get prescribed by HealthRisk AI]. I think I would still go to a doctor just to double-check because maybe it's not that trustworthy. […] So just to double-check with an actual healthcare professional.}” Adolescents preferred to have clinical decisions be informed by a human expert, as they did not solely trust the AI tool due uncertainties about its accuracy, as previously mentioned. While a few participants were open to this idea, they still generally favored confirmation by healthcare providers, as voiced by A03, “\textit{I mean, if it's advanced enough that healthcare professionals can trust [the prescription], I guess I would trust it too, but I would still prefer for a real person to look over it a couple of times at the end just to be completely sure because two opinions are always better than one.}” According to these quotes, adolescents valued the reassurance that came from having a real person review the AI’s outputs, emphasizing the importance of human supervision and the belief that two opinions-one from a healthcare provider and one from AI–are better than a single AI source.

\subsection{Personal Health Scenario One: Wellbeing AI}
The first personal health scenario featured Wellbeing AI, a personal health tracker designed to track everyday health data and provide customized insights and recommendations for achieving health goals. In the current concept, which focused on physical health, adolescents generally viewed Wellbeing AI optimistically as a tool for maintaining a healthier lifestyle. 

\subsubsection{Potential Benefits of Wellbeing AI: Supporting Personalized Lifestyle}
In particular, participants expected the AI technology to promote healthy habits and help people achieve health goals through personalized suggestions. A09 imagined seeing their progress for fitness related goals, such as exercise and nutrition, and making adjustments based on the data to improve: “\textit{I think I would like to see […] a lot of the calorie intake and a lot of my personal dietary, but I would also to see the progress that I've made or any fitness related goals so when I'm running and stuff, when I'm doing any exercises that I like, how that benefits me in a way.}” Participants particularly appreciated Wellbeing AI’s potential features of providing custom recommendations and visualizing goal progression, believing that the tool could guide individuals towards beneficial health directions. A14 highlighted this point: “\textit{I think that's pretty helpful. It would definitely make health decisions a lot easier for people or at least help them stay more informed about what to do.}” In particular to the adolescent population, A15 believed Wellbeing AI could help teenagers establish good habits at a young age: “\textit{I think it would be more helpful because it would be able to show teenagers how to be really healthy early on in their lives so that they could develop habits that would last throughout their life, so that they're mostly healthy for the entirety of their life.}”

\subsubsection{Perceived Concerns of Wellbeing AI: Inaccurate Results and Overreliance on AI}
Although the attitude towards Wellbeing AI was mostly positive, there were two main concerns raised. First, based on their experiences with existing trackers, adolescents felt Wellbeing AI could also report inaccuracies, as A07 commented, “\textit{Sometimes the Fitbit or Apple Watch miscounts your steps.}” However, they were quick to qualify their remark, stating, “\textit{But it's not really that big of a deal.}” This could be because A07 viewed their wellbeing data casually: “\textit{I just use Apple Health to track my steps or whatever. I don’t really monitor it that much.}” Adolescents perceived AI outputs related to wellbeing, such as physical activity or nutrition, as less serious for their everyday health compared to AI outputs related to medical conditions. In contrast to the previous AI technologies like ScribeBot and HealthRisk AI used in clinical settings, its perceived accuracy was less of a critical risk.

Another concern expressed by some adolescents was that relying too heavily on the AI tool could hinder development of health management skills that require critical thinking and decision-making. A12 brought up the risk of overreliance: “\textit{I feel like if you were to use this, it would make you more dependent on it because if you kept getting recommendations on meeting your goals, you wouldn’t really know what to do if you didn’t have the health tracker.}” A12 elaborated further, recognizing the potential lack of self-reflection when given the AI recommendations: “\textit{I think the problem would just be not being able to think for yourself on what you need to be doing every day or what you need to do to reach your goals.}” When simply given instructions to follow mindlessly, adolescents may lose the opportunity to think critically of their health conditions and make decisions independently to manage their health effectively. 

However, when probed about data risks, the majority of the participants were not concerned about potential data leaks for two main reasons. First, adolescents clearly viewed this type of wellbeing data, which included nutrition, activity, sleep, water, and more, as “\textit{not sensitive information}” (A08) and “\textit{shallow}” (A01). A02 questioned how one would misuse their data: “\textit{What’s someone gonna do with that information?}” Some were tolerant of this risk due to the mundaneness of this data, as A01 noted: “\textit{Even if that [wellbeing] data was leaked, I would be okay with that because it’s not anything super important.}” Second, adolescents believed they did not have personal information inside wellbeing data that could be damaging if exposed. A13 surmised that at this particular time, they don’t have any “\textit{important information that could be used against me.}” However, they further explained that the risk might grow “\textit{as I get older, it will definitely be more serious as I have an income and I have all this bank information.}” These comments indicate that adolescents were unaware of any potential negative consequences of breaches regarding wellbeing data as they considered them as trivial and low risk.

\subsubsection{Future Concept of Wellbeing AI: Valuing Direct Connections to Doctors but Rejecting Invasive Tracking}
When probed about having Wellbeing AI send their health information to their doctor upon predicting an upcoming health issue, the majority of the participants conveyed enthusiasm about the prospect of being connected to their providers. A15 even foresaw life-saving opportunities: “\textit{I think the part about it being able to detect immediate health problems and stuff and connect you with a therapist or a doctor could be really useful. And it could save a lot of people's lives.}” Having that open line of communication was viewed favorably since patients may not truly understand the extent of their everyday health situation before their hospital visit. A14 recalled volunteering at their local hospital and interviewing sepsis patients, “\textit{Whenever I interviewed them, they said they never really took their symptoms seriously at first until it got really bad. So I think if the doctors are alerted by the [AI health tracker] app, it could help prevent a lot of what I've seen happen in the hospitals where they're in really bad conditions because of how long they ignored their condition or like their symptoms.}” Here, A14 experienced first-hand the serious consequences of not understanding the gravity of one’s health status and asserted that an AI tracker could encourage patients for earlier treatment through their providers. These observations suggest that adolescents preferred professionals to be more involved in their health monitoring to assist in critical situations as they lacked the capability to understand the data correctly. However, when probed about adding sentiment analysis and facial recording features to Wellbeing AI, participants voiced their discomfort. As A04 explained, “\textit{I think the part where it can detect your emotions is kind of like, I don't know, it feels like an invasion of privacy. […] I just feel like that gets to the point where it's too involved in your life.}” Adolescents felt that emotion tracking would entail constant observation of the face, which they found overly intrusive. Similar to the ScribeBot (Section 4.1) scenario, participants felt that the type of data recording and amount of information collected were extreme. A14 voiced their discomfort about the collection of extensive identifiable data: “\textit{I guess that would feel pretty invasive if they're tracking me to my emotions and my voice. I think at that point, I would start to feel uncomfortable with having them collect all of that data.}” While adolescents perceived little to no risks with using wellbeing data, their attitudes shifted when it came to emotion and facial data, as these introduced concerns about privacy invasion and excessive data collection.

\subsection{Personal Health Scenario Two: MedHealthGPT}
The second personal health scenario presented MedHealthGPT, an AI chatbot designed for symptom evaluation, health assessment, and emotional support. The opening idea was centered around users discussing their symptoms to find possible causes for physical and mental health issues. 

\subsubsection{Potential Benefits of MedHealthGPT: Helpful for Minor Issues}
Many participants found this idea particularly helpful for discussing minor, less critical symptoms and “\textit{helping people know what's going on with them}” (A15). It was viewed to be useful in obtaining preliminary information about one’s current health situation and trying to “\textit{solve their illness especially if it’s a mild one}” (A01). Rather than being concerned, adolescents were more curious about understanding the underlying causes of minor issues, as evidenced by A14: “\textit{Maybe if I have a headache, maybe I'll just see, ‘What does the chatbot think of this?’}” However, they would prefer to go straight to their doctors for severe situations, as A14 continued: “\textit{But I think for the most part, if I had more serious symptoms, I would not turn to a chatbot, and I would probably just go to the hospital.}” Additionally, some adolescents considered that interacting with an AI chatbot could lead to more reliable results compared to just searching for symptoms on a search engine. As A16 put it, “\textit{Yeah, maybe because it converses with you and you know, has a deeper conversation than searching a one-liner on Google, it can definitely be more accurate.}” In terms of less severe cases, adolescents acknowledged the AI tool’s benefits in providing practical guidance about understanding their symptoms over a hospital visit or online searches.

\subsubsection{Perceived Concerns of MedHealthGPT: Potential Misdiagnosis Due to Complexity of Mental Health and Previous Experiences with AI}
Participants conveyed more wariness towards MedHealthGPT compared to the other scenarios, expressing two major issues. The primary issue highlighted was the potential misdiagnosis of mental health conditions. Adolescents believed that the AI chatbot lacked empathy and human experience which could lead to inaccurate analysis. A04 explained that while humans are more empathetic, “\textit{a robot just interprets what you're saying and then, like, ‘Oh, this is your diagnosis.’}” In a similar vein, A10 underscored the importance of communicating with an actual human rather than a machine: “\textit{I think regarding mental health, I think I trust talking to a real person a lot more […] I think that the human aspect is so much more important to me.}” On the whole, adolescents viewed AI systems as not fully understanding their mental health experiences and, as a result, failing to make accurate assessments. Furthermore, participants felt that information might not always be correctly conveyed through messages through text, making it difficult for MedHealthGPT to fully understand their issue and properly diagnose it. A01 detailed their opinion, "\textit{I feel like [mental states] just fluctuates so much that it'd be hard for a chatbot to tell from one conversation, or 5 minutes of talking with it, to determine whether or not you have an actual mental illness or not because a lot of the time, those mental issues are really complicated.}” Participants found it difficult to imagine an AI chatbot that could appropriately evaluate intricate mental states with limitations on human likeness and text-based conversation.

The second concern reported was that previous dissatisfying experiences with other AI technology were found to diminish trust. We noticed that adolescents who have used popular AI chatbots (e.g., ChatGPT) or search engines more frequently tended to have less trust in the health AI tools as they have experienced inaccurate results. A16 voiced their doubts about trusting MedHealthGPT’s analysis and diagnosis, comparing it to ChatGPT: “\textit{ChatGPT–it’s good in some ways but it’s not fully there in other instances. Yeah, I would not trust ChatGPT to make a diagnosis.}” Based on their observations of ChatGPT, A16 was more skeptical of MedHealthGPT. Additionally, many participants had similar experiences of searching up symptoms on Google and getting either ambiguous or alarming results such as cancer, which also lowered their trust in utilizing AI chatbot for health reasons. For instance, A14 explained, “\textit{From seeing the symptoms I've searched up on Google and the crazy like health issues they've shown me because I have one headache, I think I would just be wary of what the MedHealthGPT is suggesting, like the problems they're suggesting I have.}” In this comment, A14 generalized their disappointing interactions with Google to MedHealthGPT, adopting a cautious perspective. The careful attitude towards the AI tool stemmed from the participants’ apprehensions after reading about daunting information about their health generated from these AI products.

Due to these issues, some adolescents sought verification of the AI results by their healthcare providers, despite MedHealthGPT being a consumer-facing tool. A05 found the tool to be informative but did not see it as the primary source for healthcare information: “\textit{I don't know if I would solely rely on just using their suggestions and information, and I would probably want a second opinion on it.}” Correspondingly, A15 stated, “\textit{I would consider it as a possible diagnosis, but I would only really trust a diagnosis from a doctor.}” They considered the information from MedHealthGPT as supplementary reference for knowledge about their health situation and trusted their providers for a true diagnosis. 

\subsubsection{Future Concept of MedHealthGPT: Accessible but Not a Replacement for Human Connection}
When asked about the future possibility of MedHealthGPT as a human-like therapist, participants observed some probable benefits as well as apprehensions. Several adolescents valued the accessibility of mental health support. Many adolescents saw MedHealthGPT as a convenient resource for talking about their personal feelings and issues. A07 explained how they would use it: “\textit{I feel like that would be really helpful, especially if you're feeling really down, and you really need someone to talk to. […] I feel like having a virtual therapist is really cool.}” In this case, adolescents appreciated having the AI tool as an outlet for unloading their feelings. Others specifically mentioned that it would be more beneficial for those less willing to seek professional help. For example, it could really benefit teenagers struggling with addiction who are too embarrassed to reach out as stated by A04: “\textit{I think there's a lot of people who struggle with addiction and they don't want to admit it to their parents or a counselor. So talking to an AI robot basically, it's like they don't need to admit it to anyone. They can do therapy sessions without having to worry about someone else knowing or finding out about their addiction and get help at the same time, so I think in that way it could be helpful, especially for people my age.}” Several participants also believed the tool to be more approachable to those who “\textit{don’t want to talk to real people}” (A11). A11 elaborated, “\textit{It would be helpful in that sense. They don't have to interact with a real person, and if the chatbot can guide them through the mental health issue, then it would be good.}” These comments suggest that AI technology could potentially offer a safe avenue for adolescents to seek help, especially when they find it challenging to do so otherwise.

On the other hand, while the majority of the adolescents considered the futuristic virtual therapist MedHealthGPT as helpful, they did not believe that the tool could replace human connection. A13 argued that “\textit{at the end of the day, the AI won’t have the personal connection and personal experience that a human psychiatrist could or a mental health specialist would},” emphasizing the lack of humanness in the tool. Many preferred to consult with a human doctor. For example, A08 said, “\textit{For sure, this MedHealth can definitely help, but I don’t think it should be a complete substitute of seeking professional help. Yeah, I would still trust an actual person and human being with other people more.}” A08 believed that MedHealthGPT would not be able to “[get] to the root of the problem properly” as humans went through “many, many years of medical training.” For some, it was difficult to envision using MedHealthGPT as a virtual therapist and receiving the same support and guidance as they would from a human therapist.

\begin{table*}[ht]
\footnotesize
\resizebox{\textwidth}{!}{%
\begin{tabular}{p{4cm}p{5cm}p{5cm}}
\toprule
\textbf{Health AI Technology} & \textbf{Perceived Benefits} & \textbf{Perceived Concerns} \\ 
\midrule
ScribeBot (clinical) & - Saves time and captures information doctors might miss \newline - Enables doctors to focus more on patients & - Misinterpretation by AI \newline - Confidentiality concerns such as potential data leaks to public and to parents \newline - AI listening in on sensitive topics \newline - Excessive monitoring and data collection from camera recording \\
\\
HealthRisk AI (clinical) & - Risk score seen as personalized and trustworthy 
\newline - Helpful to learn about health and how to take preventative action & - Less accurate for mental health and severe health conditions \newline - Less accurate due to lack of personal context and human insight \newline - Lack of human doctor oversight for AI issuing prescriptions \\
\\
Wellbeing AI (personal) & - Supports lifestyle goals through personalization \newline - Promotes healthy habits early in life \newline - Direct connection to doctors & - Inaccurate tracking and outputs \newline - Overreliance leads to less critical thinking for health decisions \\
\\
MedHealthGPT (personal) & - Helpful for minor health issues \newline - Resource for guidance on practical health concerns \newline - More accessible to adolescents & - Misdiagnosis due to complexity of mental health \newline - Low trust based on previous experiences with AI \newline - Lack of human understanding \\
\bottomrule
\end{tabular}%
}
\caption{Summary of perceived benefits and concerns of health AI technologies}
\label{tab:my-table-2}
\end{table*}

\subsection{Summary}
The goal of the scenarios was to probe adolescents’ perceptions of health AI technologies across various contexts. We found that adolescents recognized benefits like improved doctor-patient interaction, personalization, risk-based preventive care, support for healthy habits, and symptom tracking. However, they raised several concerns about accuracy and data privacy, particularly for mental health and severe conditions, citing AI’s lack of personal context and human understanding. When envisioning future ideas of health AI technologies, adolescents expressed some potential advantages such as direct communication with doctors and accessibility to health-related resources. Nevertheless, most viewed them more negatively, such as the overly invasive nature of excessive tracking. They believed that these health AI technologies would not be able to replace human connection and hoped that humans would still be actively involved in any health decision-making process. We summarize these findings in Table 2.

\section{DISCUSSION}
In this section, we discuss how adolescents found health AI technologies as valuable tools for learning about their health, symptoms, and habits. However, alongside the positive sentiment, we found that adolescents have nuanced and situated concerns about using health AI that are unique to their age group. Adolescents emphasized the importance of human involvement in various ways, including enhancing the accuracy of AI technologies, actively collaborating with AI and doctors, and providing emotional support. We suggest that adolescents’ trust of health AI technologies is deeply influenced by various factors and their privacy concerns include confidentiality and excessive data collection.

\subsection{Unique Needs for Designing Health AI for Adolescents}
Echoing general public sentiment \cite{kelley_exciting_2021, beets_surveying_2023}, our findings highlight that although adolescents have mixed attitudes about health AI in both clinical and personal contexts \cite{richardson_patient_2021, young_patient_2021}, they viewed AI as a valuable resource, particularly for taking preventive measures regarding their family health history and utilizing everyday health data to promote a healthy lifestyle. Unlike adults, who viewed AI to be useful for tracking health data for future clinical visits \cite{wu_public_2023}, we found that adolescents placed emphasis on the opportunity AI provided to learn health management skills from its outputs (e.g., risk score). Given that adolescence is a critical time for gaining independence and forming health habits that often carry into adulthood \cite{frech_healthy_2012, lawrence_health_2017}, it is essential to provide the right tools to encourage learning and habit-building experiences that could remain throughout their lives. We emphasize the need for health AI technologies to create spaces for reflection and learning, rather than simply delivering passive prediction scores and recommendations \cite{mamykina_grand_2022, stromel_narrating_2024}.

AI has long been a prominent tool in education, enhancing learning through children’s TV shows \cite{xu_articial_nodate}, interactive storytelling agents \cite{zhang_mathemyths_2024}, and personalized feedback systems \cite{pankiewicz_large_nodate}. These AI-powered tools often utilize LLMs to generate custom feedback, hints, and explanations for scaffolding, or the practice of providing temporary guidance as individuals learn new concepts or skills. This is especially useful during adolescence when individuals develop more cognitive skills, such as reflection and critical thinking \cite{bonnie_promise_2019}. Given AI’s success in education, similar approaches could be applied to adolescent health.

Future health AI technologies should be designed to allow adolescent users to interact with the outputs, providing clear insights that connect their health history, behaviors, and current health status. LLM-powered systems (e.g., ChatGPT, Gemini, Llama) could be used to create resources that support learning and engagement with health information \cite{hu_grow_2024, nepal_contextual_2024}, similar to how Mollick \& Mollick (2024) provided a range of exercises and prompts for instructors to create personalized learning experiences for students \cite{mollick_instructors_2024}. LLMs could also generate simplified explanations about health information, provide real-time feedback on health behaviors, and prompt adolescents to engage in self-reflection. To ensure high quality responses from LLMs, designers can leverage Retrieval-Augmented Generation (RAG), a technique that enhances LLM accuracy by pulling information from credible external sources before generating responses \cite{gao_retrieval-augmented_2024}. Strategic integration of AI-driven feedback and interactive learning in health management could empower adolescents to make informed decisions about their well-being.

However, given participants’ concerns about overreliance on AI, the systems designed for adolescents should equip them with the right knowledge and skills for health management and provide opportunities to support adolescents’ AI literacy. AI literacy focuses on understanding core competencies that enable people “to critically understand, evaluate, and use AI systems and tools safely and effectively” \cite{mills_AI_2024, long_what_2020}. As AI becomes increasingly prevalent in everyday applications, adolescents who properly understand how AI works may be better prepared to use and apply these technologies productively. This is especially important in addressing overreliance, which occurs when users uncritically accept incorrect AI outputs due to limited knowledge, expertise, or task familiarity \cite{passi_overreliance_nodate}. By incorporating methods that explain how the health AI system functions, adolescents can take a more critical approach to understanding and utilizing the outputs. Health AI technologies should be designed specifically for adolescents to promote learning and critical thinking about health AI and to mitigate risks like overreliance.

\subsection{Human Involvement in Health AI}
Our research highlights how crucial human involvement is for adolescents regarding health AI. Adolescents valued human involvement in supporting AI’s conversational functions and health-related decision making, aligning with the Human-in-the-Loop (HITL) concept but in distinct ways. Research on HITL has typically focused on how humans can actively play a role in the development, deployment, and refinement of AI-driven systems by providing feedback \cite{benedikt_human---loop_2020, budd_survey_2021, wu_survey_2022}. Building on this idea, our findings emphasize the importance of humans in supporting AI’s conversational functions, which has been less examined. Adolescents, in particular, believed that humans are better positioned to interpret nuanced meanings in human conversations, whereas AI might struggle to accurately transcribe and understand the subtle aspects of human dialogue. Conversations about health may be particularly nuanced, especially when discussing sensitive issues, and it may be difficult to depend solely on AI to capture the essence of the dialogue. While previous studies have noted concerns about AI’s limited understanding in decision-making contexts \cite{nelson_patient_2020, young_patient_2021}, our findings suggest that these limitations extend to other scenarios, such as interpreting complex conversations during exams. Although AI ambient scribes are increasingly utilized by clinicians \cite{tierney_ambient_2024}, some technical challenges, such as difficulty in capturing non-lexical conversational sounds, have been reported \cite{van_buchem_digital_2021, tran_mm-hm_2023}. Involving humans to support conversational abilities of health AI technologies by reviewing, correcting, and annotating outputs may improve the overall accuracy \cite{budd_survey_2021}. Health AI technologies should incorporate opportunities for provider and patient to provide feedback and report errors feedback \cite{tierney_ambient_2024} while considering providers’ limited time and energy to ensure successful integration and usage.

While HITL research traditionally focuses on improving AI’s technical capabilities, our findings extend the concept, revealing that adolescents value human involvement not just for expertise but also for the emotional support that technology alone cannot provide. Past studies have reported that patients prefer a human healthcare provider’s involvement when using health AI technologies for a variety of reasons such as concerns about AI’s knowledge limitations \cite{kim_how_2024} and potential communication barriers \cite{esmaeilzadeh_patients_2021}. In addition to these reasons, our findings suggest that adolescents desired human involvement to feel a sense of safety and reassurance. Although adolescents valued the potential insights the AI could provide, they did not want to depend solely on AI, emphasizing the importance of human involvement through their doctors. Their preferences extended beyond appreciating doctors’ knowledge; adolescents also believed emotional reassurance from human presence is essential when engaging the health AI technologies, illustrating the multifaceted role humans can play in supplementing the functions of AI health technologies. This finding echoes prior HCI study on having human contact when using contact tracing technologies during public health crises \cite{lu_comparing_2021}. However, different from comparing human and technology, participants in our study desire to have humans in the loop to provide emotional assurance that health AI technology may not be able to provide and have AI systems serve as a support for healthcare providers. 

In addition to valuing technical feedback and emotional support, adolescents expressed their desire to actively participate in health decision-making through health AI technologies, highlighting their potential as key contributors in human-AI collaboration. Participants viewed health AI systems as tools for engaging with doctors, discussing preventative measures, and obtaining preliminary symptom information, suggesting how these technologies can empower adolescent patients to be more proactive in their health journey. Previous studies have identified barriers such as limited doctor-patient time, education level, and power imbalance in the patient-provider relationship as factors affecting the health decision-making process \cite{joseph-williams_knowledge_2014}. However, health AI technologies could address these issues by providing more time with the doctors, better access to health information, and greater possibilities to engage in the health decision-making process. This highlights a need to move beyond the clinician-AI dynamic and explore ways to involve doctors, AI, and adolescent patients effectively \cite{jacobs_designing_2021}. While previous studies on human-AI collaboration have mainly centered on clinicians’ workflows, we investigated adolescent patients’ perspectives on this dynamic, supporting the concept of framing AI-powered technologies as \textit{shared technologies} \cite{kim_how_2024}, where patients, alongside doctors, are considered crucial actors of human-AI collaboration. Rather than purely focusing on improving the technical capabilities of health AI technologies, adolescent patients desire to engage with the tools to become more informed and proactive in health decision-making, playing the role of another "human" in the loop. This framework of shared technology can guide the incorporation of patient-centered design into future health AI technologies, empowering adolescents to actively participate in decision-making early in life.

\subsection{Multifaceted Factors on Perceived Trust of Health AI}
Prior work has extensively focused on how transparency and high accuracy levels influence people’s perceptions on AI systems \cite{zhang_effect_2020, stephanidis_human-centered_2020, ehsan_expanding_2021, panigutti_understanding_2022, panigutti_co-design_2023}. Extending this line of work, our study found that adolescents considered trust to be a key issue for accepting health AI technologies, with factors such as complex health conditions, empathy, and previous experiences with AI systems influencing their level of trust. 

We found that adolescents express trust towards AI systems for general health conditions, they believe that complex health conditions, such as mental health or rare illnesses, are too intricate and difficult for AI to analyze. For instance, our finding suggests that adolescents value AI predictions and diagnosis for less severe physical health conditions, such common colds, allergies, or minor skin conditions. This can be beneficial, as they can utilize health AI technologies to learn how to manage these non-critical conditions independently. Moreover, we found that adolescents placed less trust for mental health prediction and diagnosis (e.g., HealthRisk AI and MedHealthGPT) compared to physical health, due to the health AI systems’ limited understanding of their personal context. While there is growing research on improving detection and diagnosis for mental health illnesses \cite{graham_artificial_2019}, many studies using machine learning techniques for children and adolescent health consider this population as a demographically homogeneous group for ages 0-18 \cite{hoodbhoy_machine_2021}. However, this may affect the accuracy of results for adolescent patients, as these models may not fully account for the diverse developmental stages and unique experiences that can impact adolescent mental health \cite{lee_adolescent_2014, gunnell_adolescent_2018}. Researchers and developers must consider ways to ethically collect and incorporate adolescents’ data into the training models used to develop health AI technologies. UNICEF, an agency of the United Nations, has published an AI policy that specifically focuses on children, offering guidance on designing child-centered AI, which also calls for the design of AI health technologies to use relevant and representative datasets to achieve high accuracy and robustness \cite{future_of_life_institute_eu_nodate}.

Empathy is another factor influencing adolescents' trust in AI for health. They feel that AI systems, lacking the ability to understand the emotional context or personal experiences, may fail to fully grasp their perspective—something that human therapists, with their emotional insight and long term relationship with patients, are better equipped to do. The desire for human empathy in technology has been reported in prior literature where chatbots often fall short to support users’ expectations \cite{brandtzaeg_chatbots_2018, jain_evaluating_2018}, likely due to their lack of human-like conversational dynamics and emotional engagement \cite{jain_evaluating_2018, liao_all_2018}. Due to the unique barriers adolescents face in accessing health services \cite{fuentes_adolescents_2018, garney_social-ecological_2021}, chatbots can be much more accessible than real therapists \cite{sweeney_can_2021, tosti_using_2024}. Integrating empathetic conversational styles could be especially beneficial for adolescents, as emotional support \cite{daher_empathic_2020, lopatovska_capturing_2022, you_beyond_2023} and self-disclosure \cite{lucas_reporting_2017, lee_i_2020} can encourage them to open up about their feelings, which has been shown to help alleviate emotional distress \cite{lee_loneliness_2016, tosti_using_2024}. Our finding aligns with prior HCI work, which contends that human characteristics like empathy can be leveraged to build trust in technology \cite{chaves_its_2019, garg_conversational_2020, ponnada_reimagining_2020}. 

Finally, we found that, surprisingly, adolescents’ previous experiences with other AI platforms negatively impacted their perceived trust, creating a barrier to accepting the proposed health AI systems. Reflecting on their past experiences with popular AI chatbots (e.g., ChatGPT) and AI-generated summaries (e.g., Google’s AI Overview), adolescents noted that while these technologies provide convenience in searching for information, they occasionally produce inaccurate or vague results. As a consequence, these disappointing instances led to worries about trusting health AI systems and their output (e.g., risk scores and mental health diagnosis), since they believed such systems should have a high level of accuracy. Prior work has indicated that past experiences can shape users’ mental models of AI \cite{nourani_anchoring_2021, nourani_importance_2022, kahr_understanding_2024}. For instance, Nourani et al. reported that short-term and long-term past experiences with AI shape users’ mental models, which calibrates their level of trust and reliance on the outputs \cite{nourani_importance_2022}. Adolescents’ dissatisfying experiences with AI in educational contexts qualified their expectations of high accuracy and lowered their trust in health AI. These biases, shaped by adolescents’ impressions, can influence user behavior and impact the effectiveness of human-AI collaboration, potentially affecting healthcare decision-making \cite{nourani_importance_2022}. This is important to consider when designing for adolescents, as both under- or over-relying on AI could lead to suboptimal health outcomes \cite{bansal_does_2021, bucinca_proxy_2020}.

\subsection{Situated Health AI Data Privacy Risks}
Previous research has identified disability \cite{ahmed_privacy_2015, akter_i_nodate}, culture \cite{ahmed_digital_2017, jack_privacy_2019}, socioeconomic background \cite{sambasivan_privacy_nodate, al-ameen_we_2020}, and gender \cite{fernandez_i_2019, lerner_privacy_2020} as factors that influence people’s perception of privacy, emphasizing the understanding that privacy is contextual and could be significantly influenced by one’s characteristics. For instance, people with visual impairments experience unique privacy risks as they may need to require more assistance from strangers for tasks involving personal information, such as asking bank employees to fill out financial documents \cite{ahmed_digital_2017}. Similarly, our findings suggest that adolescents also have unique concerns distinct from general privacy and data issues. First, adolescents voiced confidentiality issues particularly regarding their parents, reflecting a layer of privacy concerns unique to this age group. Second, we found that adolescents’ perception of privacy is affected by certain characteristics of the data, such as the sensitivity of their medical and health data or the additional types of recording used by AI to facilitate medical practices and care. 

Our findings indicate that adolescents were specifically nervous about sharing sensitive information, such as mental health issues, in front of the ScribeBot as they were unsure whether the AI would relay this information to their parents. This suggests that the perception of confidentiality may affect their desire to disclose and discuss sensitive information in the presence of health AI technologies. Confidentiality issues can be a serious barrier for adolescents in discussing sensitive health concerns and seeking health care \cite{ford_confidential_2004, carlisle_concerns_2006, thrall_confidentiality_2000}, and the addition of AI presence could potentially impede honesty during the medical exam. This is an interesting experience as adolescents are still under the care of their parents or guardians, but have expressed wanting more independence in managing their own health \cite{toscos_best_2012, pina_personal_2017}. As adolescents have expressed a desire to have the ability to control what information to share with \cite{bergman_teen_2008}, it may be unclear how health AI technologies will safeguard sensitive data from potential access and guarantee perceived safety for adolescents. Part of the challenge lies in the granularity of the data being shared, which can be interpreted in various ways \cite{caine_patients_2013}, and in the complexity of electronic health record (EHR) access in adolescence. Adolescent and parental access to EHR is complicated as minors can request to keep certain health information confidential from their parents. Now, the presence of AI may add another layer of challenges as adolescents may be concerned that the AI system would somehow affect confidentiality. Without a clear data transparency of the AI systems, adolescents may hesitate to share sensitive information with their medical providers. While future health AI systems should handle adolescent and parental access to confidential data responsibly, providers must focus on safe disclosure practices while using these systems to create a supportive environment for adolescents.

Beyond confidentiality issues, we found that adolescents had their own perception of what sensitive data entails, affecting their privacy concerns. While often treated as a whole, a prior study found that the data within a medical record varies in sensitivity and may require different levels of consent \cite{bourgeois_whose_2008}. Other studies have noted the complexities and discrepancies of users’ perceptions about their privacy across the context of clinical care and personal health management \cite{gupta_individuals_2016, cherif_personal_2021, tran_patients_2019, gabriele_understanding_2020}. For example, individuals exhibited lower privacy concerns regarding data collected from fitness trackers as they believed the threats to be very unlikely to occur \cite{gabriele_understanding_2020}, but in a study by Tran et al., only 20\% of participants considered the benefits of the wearable to outweigh the dangers \cite{tran_patients_2019}. However, in the clinical context, the perceived benefits positively impacted privacy concerns \cite{gupta_individuals_2016, cherif_personal_2021}. While these studies noted the differences, they were conducted in different contexts with different research goals. By comparing users’ perceptions towards different forms of data in one study, we were able to conduct a comparative analysis that highlights how adolescents, in particular, exhibit different perceptions regarding the broad spectrum of medical and personal health data. For example, conversations during general medical consultations were considered non-sensitive while contentious topics such as mental health and abuse were viewed as requiring greater privacy. 

Outside the medical setting, adolescents also viewed wellbeing data as non-sensitive, consequently expressing even less concern about potential risks. Our findings confirmed the prior research that users have varied level’s privacy concerns when data carries different sensitivity. Additionally, in both clinical and personal health contexts, we found adolescents were willing to accept these risks in exchange for the perceived benefits of the health AI technologies, demonstrating the trade-off between the perceived benefits and risks in different medical and personal health situations. However, it is also highly likely that adolescents are unaware of many potential privacy risks, (e.g., bias, inferences drawn data) due to a general lack of privacy knowledge among people \cite{velykoivanenko_are_2021} and the unique set of privacy risks introduced by AI technologies \cite{lee_deepfakes_2024}.

Additionally, we found that adolescents’ privacy concerns also vary based on the type of additional data being recorded by AI. They were accepting of collecting voice data in conversations but not of face and body video recordings, noting that facial and body recordings were too excessive and unnecessary. However, adolescents’ concerns focused more on the feeling of privacy invasion rather than potential data misuse, as they admitted uncertainty about how their data could be misused. Prior work has indicated that people generally either have low concerns \cite{ajana_personal_2020, gabriele_understanding_2020} or a lack of knowledge regarding data privacy issues \cite{chowdhury_privacy_2018, velykoivanenko_are_2021}. The vast data requirements of AI heighten these risks, as it reduces control over what is collected. Adolescents’ differing perceptions of type of data recording also demonstrates that although they have general concerns about sharing their medical record data, they view excessive data collection for both clinical and personal health purposes as a more serious privacy invasion.

Together, our study suggests that adolescents are willing to share their existing medical and health data, especially when trading it for personal benefits, unless the data is viewed highly sensitive. This differing privacy concerns towards the sensitivity, and types of recording provide additional insights into adolescents' privacy consideration, highlighting the need to consider the unique needs of this age group. Future health AI systems should account for the nuances of data sharing, balancing the perceived benefits, data sensitivity, and the nature of additional recording in facilitating medical treatment and healthcare. These understandings present opportunities to build health AI technologies that protect adolescents from certain privacy risks, foster learning in managing their health, and empower their agency in healthcare.

\subsection{Limitations}
This study has several limitations. The majority of participants were recruited from two summer programs hosted by an academic research institution in the United States. As a result, we acknowledge that our findings are limited to the perspectives of the participants who have attended the summer programs and could be subject to selection bias. Moreover, most of our participants were 16-17 years old (two participants were 13), so our findings may not fully represent the views of younger adolescents, who could hold different beliefs \cite{nguyen_examining_2022}. Additionally, with only two male participants, our findings are more reflective of female adolescents’ perspectives. Finally, these participants did not disclose having any severe health conditions, adolescents with ongoing health issues may perceive AI health technologies differently, as they are more actively engaged in managing their health \cite{huh_collaborative_2012, james_chronic_2023, ankrah_plan_2023, su_creating_2024}. To address these limitations, future work could focus on a particular age group or health condition. Younger children and adolescents with chronic illnesses may hold different beliefs about health AI as the former may have less opportunity to interact with health-related technologies while the latter may use health technologies more frequently. Furthermore, we acknowledge that race and ethnicity could significantly affect perception, usage, and beliefs towards health AI as there are concerns about racial bias \cite{obermeyer_dissecting_2019}. While our participants did not explicitly mention race and ethnicity, these factors are critical, and we suggest that future research with adolescents investigate how race and ethnicity may shape their perception of health AI.

\section{Conclusion}
In this paper, we present a comprehensive understanding of adolescents’ perceptions of health AI. Using Design Fiction, we interviewed 16 adolescents to explore their perceived benefits and risks regarding health AI technologies. Our findings reveal that adolescents’ understanding of health AI is unique to their age group, and their concerns bring additional nuance to perceptions of trust and privacy in health AI. These insights underscore the importance of considering adolescent-specific needs, human involvement, and a multitude of factors that affect trust and privacy to inform future design and implementation of health AI technologies. We suggest that incorporating these insights could foster learning, encourage active engagement in health decision-making, ensure safe data privacy practices, and improve patient outcomes for adolescents, helping to realize the full potential of health AI.

\begin{acks}
We would like to thank the participants for their insights and time. We also appreciate the clinicians at UCI School of Medicine for their role in recruitment and scenario development. Further appreciation to Daniel Epstein and Elena Agapie for leading the CHI Bootcamp. This study was partially supported by grants from the UC Irvine Donald Bren School of Information and Computer Sciences Academic Senate Council on Research Computing and Libraries (CORCL) and the National Science Foundation (NSF) grant number 2211923. 
\end{acks}

\bibliographystyle{ACM-Reference-Format}
\bibliography{references}










\end{document}